\documentclass{article}
\begin{document}
\vbadness = 100000
\hbadness = 100000
\title{Comment on ``Electromagnetic mass differences of SU(3) baryons within a chiral soliton model"}
\author{Jerrold Franklin\footnote{Internet address:
Jerry.F@TEMPLE.EDU}\\
Department of Physics\\
Temple University, Philadelphia, PA 19122-6082}
\maketitle
\begin{abstract}
In a recent letter\cite{ykp}, several electromagnetic mass difference formulae for baryons were presented.
However, because the derivation did not include important colormagnetic terms, the mass relations do not correctly give isospin mass splittings for the baryons.  Correct mass formulae were published some time ago in a model independent approach that was more general and correct than the approach in this letter.  In this Comment, the errors in the letter are pointed out and  some correct formulae presented.
\end{abstract}

In a recent letter(YKP)\cite{ykp}, electromagnetic mass differences of SU(3) baryons were calculated using a 
``model-independent approach" within a chiral soliton model.  In this Comment we point out that the formulae in the paper do not represent 
the complete isospin breaking mass differences.
By limiting the calculation to purely electromagnetic interactions, the letter leaves out important  QCD hyperfine  (or ``colormagnetic") mass splitting, 
The QCD splitting was first introduced in 1975\cite{dgg}, and shown to be comparable in size to the electromagnetic splitting in 1982\cite{jl}. 
The importance of the QCD splitting was well known at the time, but seems to have been forgotten over the years.  Because of the omission of these QCD terms, most of the mass relations in YKP do not represent the full isotopic splitting.  

Correct versions of mass relations for baryons in the quark model were derived in the 1960's\cite{frt,rss,jf}.  Even though these mass relations predated the introduction of the
QCD terms, they correctly included the QCD effect because of the model independence of the relations.
It was shown in Refs.\ \cite{frt} and \cite{rss} that there are nine baryon mass relations, generally in agreement with experiment, that follow for any dynamical quark-quark interaction for ground state baryons composed of three quarks in the SU(3) octet and decuplet states of SU(6).  The restriction to SU(6) states, or any group symmetry, was removed in Ref. \cite{jf}, using only the spin dependence of the ground state baryons.  It was shown there that, with antisymmetry in another degree of freedom
(``hidden spin" or ``color"), there were eighteen ``ground state" baryons (that is, three quark baryons with no orbital angular momentum), with eight having spin 1/2 and ten having spin 3/2.  The only symmetry needed was the angular momentum addition of three spin 1/2 quarks.  This can be summarized by the following statement:
\begin{quote}
{\em The eighteen ground state baryons are composed of spin one half quarks with charges +2/3, -1/3, -1/3, whose three quark wave function is antisymmetric in the color degree of freedom, and has no orbital angular momentum.}
\end{quote}
  Notice that no mention need be made of any internal group symmetry, not SU(6), SU(3), or even SU(2) isospin.  
The resulting eighteen ground state baryons are given in Eqs.\ (1) and (2) of Ref.\ \cite{jf}, and are shown to be the usual octet and decuplet baryons (but with no symmetrization of the wave functions).   With the additional assumption of only one-body and two-body quark interaction energy contributions, the nine mass relations follow for the ground state baryons.  Any calculation that includes these minimal assumptions should be consistent with these nine mass relations, independently of the specific quark-quark dynamics.
Mass relations have subsequently been derived for charmed\cite{charm} and heavier\cite{heavy} baryons, which are generally in agreement with experiment.

We now look at some of the mass relations derived in YKP.
Equation (21)  is (using the particle symbols for their masses)
\begin{equation}
p-n=(\Sigma^+ -\Sigma^-)-(\Xi^0-\Xi^-).
\end{equation}
This is the well known Coleman-Glashow (C-G) relation, which is correct, but its interpretation is wrong.  YKP states
``Note that even though we consider the EM corrections, the C-G relation is still preserved."
Yet the C-G relation was originally derived in much the same way as in YKP,  assuming group properties of purely electromagnetic interactions.
In fact the C-G relation does not need the restriction to electromagnetic interactions in its derivation.  It is shown in Ref.\ \cite{jf} that it follows from the minimal assumptions stated above.

Equation (22) of YKP consists of two mass relations that are not given in Refs.\ \cite{frt}-\cite{jf}.  Although YKP states ``These two mass relations are
well satisfied with the experiment data.", this is not the case.  The left and right sides in Eq.\ (22) differ by 27 MeV, which is about what it would be without their electromagnetic corrections.  In looking at the baryon combinations in Eq.\ (22), it can be seen that the exclusion of important colormagnetic terms invalidate the mass relations.
Their equation (23), which is the sum of the two mass relations in Eq.\ (22), can be written as
\begin{eqnarray}
2({\overline N} +{\overline \Xi})-3\Lambda^0-{\overline \Sigma}&=&\frac{2}{3}(\Sigma^++\Sigma^--2\Sigma_0),\\
(-26)&&(1.6)\nonumber
\end{eqnarray}
where the bar over the symbol denotes an average over charge states, and we have put the experimental energy in MeV in parenthesis.
This is close to Eq.\ (24) of Ref.\ {\cite{jf}, which is
\begin{eqnarray}
2({\overline N} +{\overline \Xi})-3\Lambda^0-{\overline \Sigma}&=&{\overline \Delta}+\Omega^- -{\overline \Sigma^*}-{\overline \Xi^*}.\\
(-26)&&(-14)\nonumber
\end{eqnarray}

Each of these equations represents a correction to the well known Gell-Mann Okubo formula\cite{gmo} which has the right hand side equal to zero.
The reader can judge which formula is more reasonable.
Equation (2) is so far off because the left hand side includes colormagnetic interaction terms, while the right hand side is purely electromagnetic.
Equation (3) has colormagnetic terms on each side.
The residual error in Eq.\ (3) can be attributed to a small difference in the spatial wave functions of the spin 3/2 and spin 1/2 baryons.
The other formula in Ref.\ \cite{jf} relating spin 3/2 to spin 1/2 baryons is off by 25 MeV.

There are four mass relations in YKP for the spin 3/2 baryons, none of which are given in Ref.\ \cite{jf}.  Three of them are for $\Delta$ charge states, and cannot be reasonably compared to experiment because of experimental uncertainty.  Their relation 
\begin{eqnarray}
\Sigma^{*0}-\Sigma^{*-}&=&\Xi^{*0}-\Xi^{*-}\\
(-3.5\pm1.1)&&(-3.8\pm.7)\nonumber
\end{eqnarray}
agrees with experiment, although the experimental uncertainty is large.
However,  while the electromagnetic interaction terms are the same on each side of the equation, the colormagnetic terms are different.
There are two isospin splitting mass relations in Ref.\ \cite{jf} that relate decuplet masses to octet masses, but they involve more complicated linear combinations.  
They are satisfied within large experimental errors.  There are also three mass relations between decuplet baryons in Ref.\ \cite{jf}, but they all involve $\Delta$ charge states, and can't be tested.

In summary, the mass relations and presumably other results in YKP are flawed because they do not include colormagnetic interaction terms.


\begin{thebibliography}{9}
\bibitem{ykp}G-S. Yang, H-C. Kim, M. V. Polyakov,  Phys. Lett. B {\bf 695} (2010) 214.
\bibitem{dgg}A. De Rujula, H. Georgi, and S. L. Glashow, Phys. Rev. D {\bf 12} (1975) 147.
\bibitem{jl}J. Franklin and D. B. Lichtenberg, Phys. Rev. D {\bf 25} (1982) 1997.
\bibitem{frt}P. Federman, H. Rubenstein, I. Talmi, Phys. Lett. {\bf 22} (1966) 208.
\bibitem{rss}H. Rubenstein, F. Scheck, and R. H. Socolow, Phys. Rev. {\bf 154} (1967) 1608
\bibitem{jf}J. Franklin, Phys. Rev. {\bf 172} (1968) 1807.  The ``hidden spin" in the title of this paper was an early (precolor) name for an extra degree of freedom in which the three quark wave function of a baryon was completely antisymmetric.
\bibitem{charm}J.~Franklin, Phys.\ Rev.\  D {\bf 53} (1996) 564.
\bibitem{heavy}J.~Franklin, arXiv:0811.2143 [hep-ph].
\bibitem{cg}S. Coleman and S.L. Glashow, Phys. Rev. Lett. {\bf 6} (1961) 423.
\bibitem{gmo}M. Gell-Mann, Phys. Rev. {\bf 125} (1962) 1067; S. Okubo, Prog. Theor. Phys. {\bf 27} (1962) 949.


\end{thebibliography}
\end{document}